# Flexural Mie Resonances: Localized Surface Platonic Modes

M. Farhat[1,*], S. Guenneau[2], P.-Y. Chen[3], K. N. Salama[1], and H. Bağcı[1]

[1]*Division of Computer, Electrical, and Mathematical Sciences and Engineering,*

*King Abdullah University of Science and Technology (KAUST)*

*Thuwal 23955-69100, Saudi Arabia,*

[2]*Aix-Marseille Université, CNRS, Centrale Marseille, Institut Fresnel,*

*Campus universitaire de Saint-Jérôme, 13013 Marseille, France,*

[3]*Department of Electrical and Computer Engineering, Wayne State University,*

*Detroit, Michigan 48202, USA.*

[*]mohamed.d.farhat@gmail.com

Surface plasmons polaritons were thought to exist only in metals near their plasma frequencies. The concept of spoof plasmons extended the realms of plasmonics to domains such as radio frequencies, magnetism, or even acoustic waves. Here, we introduce the concept of localized surface platonic modes (SPMs). We demonstrate that they can be generated on a two-dimensional clamped (or stress-free) cylindrical surface, in a thin elastic plate, with subwavelength corrugations under excitation by an incident flexural plane wave. Our results show that the corrugated rigid surface is elastically equivalent to a cylindrical scatterer with negatively uniform and dispersive flexural rigidity. This, indeed, suggests that *plasmonic-like* platonic materials can be engineered with potential applications in various areas including earthquake sensing, or elastic imaging and cloaking.

**Introduction.** In the past few years, there has been an increasing interest in designing acoustic and elastic metamaterials [1]. This followed the demonstration by Sheng *et al.* of the existence of a dynamic mass density and bulk modulus that can be obtained using locally resonant sonic materials [2]. However, unlike plasmonic metamaterials that rely on (localized or propagating) surface plasmons polaritons (SPPs) to generate the desired dispersion characteristics, acoustic and elastic metamaterials almost exclusively rely on the geometrical properties of their meta-atoms. This is due to the fact that surface plasmons do not exist naturally in elastodynamics [3]-[5]. As a result, design of mechanical metamaterials has made use of mostly phononic (or platonic) crystals [6],[7] counterpart of photonic crystals, and resonant cavities or pipes [8]-[13]. This hindered the development of applications of elastic and acoustic metamaterials that could benefit from elastic *plasmonic-like* features.

Pendry *et al.* proposed back in 2004 the counterpart of SPPs for RF waves (radiofrequency spectrum) where all metals behave like perfect electric conductors (PEC) [14], with no or little penetration of electromagnetic fields inside the structures [15]. The idea consisted in introducing periodic subwavelength corrugations in the PEC so that propagating modes are sustained, and in the long wavelength regime the equivalent material behaves as a Drude metal with a plasma frequency solely depending on the geometry of the corrugations. This is somewhat related to surface plasmons in diffraction gratings and their inner relation to the Wood's anomaly [16]-[18]. Soon after, the experimental validation of this concept followed in the RF regime [19] and many studies further demonstrated the importance of the obtained so-called *spoof* plasmons, that behave effectively in a similar manner to SPPs [20]-[23]. Localized spoof plasmons have been also demonstrated for RF waves and magnetism with many intriguing features such as sensing and field enhancement [24],[25]. In the same vein, subwavelength corrugations of rigid acoustic structures (equivalent of PEC) can generate the so-called surface acoustic waves (SAWs), the counterpart of spoof plasmons [3],[26]-[28]. SAWs behave in many ways as spoof plasmons (and therefore to SPPs), e.g. they have a dispersion relation that lies outside the acoustic cone and can be used for sub-resolution acoustic imaging or sensing. The obtained acoustic metamaterial possesses negative effective density that is responsible for these surface waves [29]-[32]. Localized SAWs have been also developed using similar structures, i.e. corrugated rigid cylinders of spheres [5].

In this work, a fourth order biharmonic wave equation with appropriate boundary conditions, which is pertinent for the scattering of flexural waves, is derived from the generalized elasticity theory [33],[35] and is used in designing an anisotropic metamaterial that operates in thin plates [36]-[41]. We then, first analyze the scattering properties of a cylindrical inhomogeneity with tailored flexural rigidity and density and show its intriguing features, when these parameters take negative values. We then go one step further by discussing the response of the platonic structure shown in Fig. 3(a), consisting of a corrugated cylinder in a thin plate with rigid grooves, in the presence of a plane wave elastic excitation (harmonic vibration of the plate in the vertical $z$-direction). It is assumed that the out-of-plane dimension of the plate is negligible compared to its in-plane dimensions [35]. We show that in the quasistatic limit, *i.e.* for $\beta_0 a \ll 1$ where $\beta_0$ is the flexural wave wavenumber and $a$ is the size of the scatterer, the scattering is dominated by the zero order multipole, unlike in the electrodynamics case where the first significant order is the dipolar one. This is not the only difference between the two scenarios: the fourth order biharmonic partial differential equation, which typically describes the propagation of bending waves in ultra thin plates, is not equivalent to the vector/scalar wave equations that describe electromagnetic or acoustic wave propagation. Consequently, one can anticipate that new Mie resonant modes and new relevant physics will be introduced following this route.

**Set up of governing biharmonic equation.** Let us start by invoking the Kirchhoff approximation for thin plates (when the flexural wavenumber is small in comparison with the plate thickness i.e. $\beta_0 h \ll 1$) i.e. by assuming that shear deformation and rotary inertia are both negligible in comparison to pure bending [33],[34]. For the sake of convenience, we start with Cartesian coordinates $(x,y,z)$, with $z$ the vertical component. The general biharmonic equation for plates of variable thickness $h(x,y)$ along $z$ is, therefore [36]

$$\rho h \frac{\partial^2 \zeta}{\partial t^2} + D\Delta^2\zeta + \Delta D \Delta\zeta + 2\frac{\partial D}{\partial x}\frac{\partial}{\partial x}\Delta\zeta + 2\frac{\partial D}{\partial y}\frac{\partial}{\partial y}\Delta\zeta - (1-\nu) \times$$
$$(\frac{\partial^2 D}{\partial x^2}\frac{\partial^2 \zeta}{\partial y^2} - 2\frac{\partial^2 D}{\partial x \partial y}\frac{\partial^2 \zeta}{\partial x \partial y} + \frac{\partial^2 D}{\partial y^2}\frac{\partial^2 \zeta}{\partial x^2}) = 0, \quad (1)$$

where $D = Eh^3/[12(1-\nu^2)]$ is the flexural plate rigidity, $E$ the Young modulus, $\nu$ the Poisson's ratio, $\rho$ the density, and $\zeta(x,y)$ the vertical displacement. However, one notes that $D\Delta^2\zeta \gg \Delta D \Delta\zeta$ as the vertical displacement varies on the same order as the flexural wave wavelength, and the variation of plate rigidity is slow enough. Similarly, $D\Delta^2\zeta$ dominates all remaining terms of the equation. Assuming a time harmonic dependence, we end up with the Kirchhoff-Love equation for thin plates:

$$\Delta^2\zeta - \beta_0^4\zeta = 0, \quad (2)$$

with $\beta_0^4 = \omega^2 \rho h / D$ the flexural wavenumber, $\omega$ the wave angular frequency. Alternatively, let us start with the Kirchhoff-Love approximation for thin plates of constant thickness but spatially varying elastic parameters. We use polar coordinates as we have in mind the design of Figs. 1(a) and 3(a) for the homogenization algorithm. We still assume that shear deformation and rotary inertia are both negligible in comparison to pure bending [34]. The vertical displacement of the plate $\zeta(r,\phi)$ obeys therefore

$$\nabla\cdot\left\{E^{1/2}\nabla\left[(\rho/\rho_0)^{-1/2}\nabla\cdot\left(E^{1/2}\nabla\zeta\right)\right]\right\} - (\rho/\rho_0)^{1/2}\beta_0^4\zeta = 0. \quad (3)$$

**Condition of resonant scattering from cylindrical objects.** We consider, first, as schematized in Fig. 1(a), an incident plane wave propagating in the $x$-direction and impinging on a cylindrical inhomogeneity located in a thin plate obeying Eq. (3). The displacement field associated with this plane wave can be expressed as $\zeta^{\text{inc}}(r,\phi) = e^{i\beta_0 r \cos\phi}$. $\zeta^{\text{inc}}$ can equivalently be expanded in terms of Bessel functions, i.e. $\zeta^{\text{inc}}(r,\phi) = \sum_{l=0}^{\infty} \varepsilon_l i^l J_l(\beta_0 r)\cos(l\phi)$, where $\varepsilon_0 = 1$ and $\varepsilon_{l \geq 1} = 2$. On the other hand, the scattered field $\zeta^{\text{scat}}(r,\phi)$ from the structure shown in Fig. 1(a) must satisfy the radiation condition at $r \to \infty$, and therefore it is represented in terms of cylindrical Hankel functions of the first kind and modified Bessel functions as,

$$\zeta^{\text{scat}}(r > a, \phi) = \sum_{l=0}^{\infty} i^l \left[ A_l H_l^{(1)}(\beta_0 r) + B_l K_l(\beta_0 r) \right] \cos(l\phi). \quad (4)$$

Inside the scatterer, and imposing that the field is finite at every point, the displacement field for $r \leq a$,

$$\zeta^{\text{in}}(r,\phi) = \sum_{l=0}^{\infty} i^l \left[ C_l J_l(\beta_{in} r) + C_l' I_l(\beta_{in} r) \right] \cos(l\phi). \quad (5)$$

The four unknown coefficients $A_l, B_l, C_l, C_l'$ are obtained by applying boundary conditions at $r = a$ requiring that $\zeta$, its normal derivative $\partial_r \zeta$, and the bending momentum

$$M_r = -D_i \left[ \frac{\partial^2 \zeta}{\partial r^2} + \nu \left( \frac{1}{r} \frac{\partial \zeta}{\partial r} + \frac{1}{r^2} \frac{\partial^2 \zeta}{\partial \phi^2} \right) \right], \quad (6)$$

and the generalized Kirchhoff stress associated with the $r$-direction,

$$V_r = -D_i \frac{\partial(\Delta \zeta)}{\partial r} - D_i (1-\nu) \frac{1}{r^2} \frac{\partial}{\partial \phi} \left( \frac{\partial^2 \zeta}{\partial r \partial \phi} - \frac{1}{r} \frac{\partial \zeta}{\partial \phi} \right), \quad (7)$$

are all continuous across the external boundary, with $D_i$ denoting the relative flexural rigidity in the appropriate domain. This results in the following linear system (for each order $l$)

$$\begin{bmatrix} H_l^{(1)}(\beta_0 a) & K_l(\beta_0 a) & -J_l(\beta_{in} a) & -I_l(\beta_{in} a) \\ \beta_0 H_l^{(1)\prime}(\beta_0 a) & \beta_0 K_l'(\beta_0 a) & -\beta_{in} J_l'(\beta_{in} a) & -\beta_{in} I_l'(\beta_{in} a) \\ S_H(\beta_0 a) & S_K(\beta_0 a) & -S_J(\beta_{in} a) & -S_I(\beta_{in} a) \\ T_H(\beta_0 a) & T_K(\beta_0 a) & -T_J(\beta_{in} a) & -T_I(\beta_{in} a) \end{bmatrix} \begin{bmatrix} A_l \\ B_l \\ C_l \\ D_l \end{bmatrix} = - \begin{bmatrix} J_l(\beta_0 a) \\ \beta_0 J_l'(\beta_0 a) \\ S_J(\beta_0 a) \\ T_J(\beta_0 a) \end{bmatrix}. \quad (8)$$

where

$$\begin{cases} S_Z(\beta_i r) = D_i [l^2(1-\nu) \mp (\beta_i r)^2] Z_l(\beta_i r) - D_i (1-\nu) \beta_i r Z_l'(\beta_i r) \\ T_Z(\beta_i r) = D_i [n^2(1-\nu)] Z_l(\beta_i r) - D_i [n^2(1-\nu) \mp (\beta_i r)^2] \beta_i r Z_l'(\beta_i r). \end{cases} \quad (9)$$

$Z_n$ are different Bessel/Hankel functions and the upper (lower) signs in the above functions $S_Z$ and $T_Z$ are respectively applied to $J_l(\bullet), Y_l(\bullet), H_l^{(1)}(\bullet) \left[ K_l(\bullet), I_l(\bullet) \right]$.

$\beta_i$ and $D_i$ denote, respectively, the flexural wavenumber and the relative flexural rigidity in the appropriate domain (free space or inside the object). In the far field, only the part of the field involving scattering coefficient $A_l$ contribute whereas the part involving coefficients $B_l$ is evanescent, since the modified Hankel functions $K_l(\beta_0 r)$ decay exponentially when one moves away from the object. In addition, if one considers small scatterers in comparison to the wavelength (the condition $\beta_0 a \ll 1$ is valid), only few multipolar orders contribute to the total scattering, since the scattering coefficients $A_l$ scale as follows

$$A_0 = -i\frac{\pi}{8}\frac{D_{in}(2-\rho_{in})-[(2-\rho_{in})v+\rho_{in}]/(1+v)}{D_{in}+(1-v)/(1+v)}(\beta_0 a)^2 + O((\beta_0 a)^3),$$

$$A_1 = -i\frac{\pi}{64}\frac{D_{in}(2-\rho_{in})-[\rho_{in}+2+(2-\rho_{in})v]/(3+v)}{D_{in}+(1-v)/(3+v)}(\beta_0 a)^4 + O((\beta_0 a)^5), \quad (10)$$

$$A_2 = -i\frac{\pi}{8}\frac{D_{in}-1}{D_{in}+(3+v)/(1-v)}(\beta_0 a)^2 + O((\beta_0 a)^3),$$

$$A_l = o((\beta_0 a)^2),\ l = 3,4...$$

Here we make use of the Landau notations $o$ and $O$. One can see immediately that in sharp contrast to electromagnetic and acoustic waves, the second order scattering $A_2$ is dominant in comparison to the first order $A_1$ and of same order as the fundamental order $A_0$. This behavior was already shown in Refs [35],[37],[38] for holes or cylindrical scatterers in the low frequency regime. The dominant scattering from the cylinders is thus given by

$$A_{tot} \approx A_0 + A_2 \approx -i\frac{\pi}{8}(D_{in}-1)\left[\frac{h_0(D_{in},\rho_{in})}{D_{in}+f_0(v)} + \frac{1}{D_{in}+f_2(v)}\right](\beta_0 a)^2, \quad (11)$$

with $f_0(v) = (1-v)/(1+v)$ and $f_2(v) = (3+v)/(1-v)$ two positive functions of the Poisson ratio $v$ that satisfies the condition $-1 < v \leq 0.5$ [33], and $h_0$ is a function of $\rho_{in}$ and $D_{in}$. In the case of a relative density $\rho_{in} = 1$, i.e. only the flexural rigidity is varying, one has $h_0 = 1$.

In the framework of elasticity, an equivalent to the optical theorem was derived to relate forward scattering to the total scattering cross section (SCS) [35]. The far field scattering amplitude (or equivalently the differential scattering cross section) $g(\phi) = \sqrt{2r}e^{-i(\beta_0 r - \pi/4)}\lim_{r\to\infty}\zeta^{scat}(r,\phi)$ is a measure of the visibility of the object in the specific direction $\phi$. The total scattering cross section is the integral of $g(\phi)$ over all angles, i.e. $\sigma^{scat} = \int_0^{2\pi} d\phi g(\phi)$. It may be expressed thus in terms of the scattering amplitudes (or coefficients) as

$$\sigma^{scat} = \frac{4}{\beta_0} \sum_{l=0}^{\infty} \varepsilon_l |A_l|^2. \tag{12}$$

Note here that coefficients $B_l$ were neglected, since in the far field the modified Bessel and Hankel functions vanish, so only the coefficients $A_l$ contribute to the far fields. Generally speaking, the possibility for an observer to detect any object in the far field is determined by the amplitude of $\sigma^{scat}$. For example, in cloaking applications [37], minimizing $\sigma^{scat}$ would lead to the undetectability of the object, irrespective of the observer position (in the far field). In this work, we are interested in the Mie resonances of elastic objects [see Fig. 1(a)] that correspond to the multipoles ($l=0,1$ represent for example the monopole and dipole modes, respectively). These correspond to local maxima of $\sigma^{scat}$. Figure 1(b) shows the contours (in logarithmic scale) of the normalized scattering cross section $\sigma^{scat}/a$ given in Eq. (12) varying the relative density $\rho_{in}$ of the obstacle and its relative flexural rigidity $D_{in}$ for the normalized flexural wavenumber $\beta_0 a = 0.1$, i.e. satisfying the quasistatic condition ($\beta_0 a \ll 1$). In this scenario, and as shown in Eq. (10), $\sigma^{scat}$ is dominated by the monopole term and the second order scattering coefficient ($A_0$ and $A_2$). The plot given in Fig. 1(b) confirms that the SCS presents two resonances (dark red color) corresponding to negative values of $D_{in}$ of -0.54 and -4.71. The dependence of the SCS on $\rho_{in}$ is however less pronounced and there are no resonances associated with the density in the whole range (negative and positive values). This is in agreement with the quasistatic expansions in Eqs. (10)-(11) that give the resonance conditions of an elastically small cylindrical scatterer ($\beta_0 a \ll 1$). The different scattering coefficients are plotted in logarithmic scale [Fig. 1(c)] for a fixed value of the density $\rho_{in} = 1.5$ and the same wavenumber [corresponding to the dashed white curve in Fig. 1(b)], versus the flexural rigidity. $A_0$ and $A_2$ are clearly orders of magnitudes larger than the remaining scattering coefficients (including $A_1$). Additionally, one can clearly see that the first resonance of $\sigma^{scat}$ [Fig. 1(b)] corresponds to the resonance of $A_0$ and more precisely to $D_{in} = -(1-v)/(1+v)$, whereas the second resonance corresponds to the resonance of $A_2$ and more precisely to $D_{in} = -(3+v)/(1-v)$. These two relations are the counterparts of the Fröhlich condition associated with the electromagnetic dipole surface plasmon [42].

In Fig. 2, we consider the example of a cylindrical scatterer with flexural rigidity described by the following dispersion function

$$D_{in} = D_0 - \beta_p^4 / [\beta_0^2(\beta_0^2 + i\gamma_b)], \qquad (13)$$

with $D_0 = 1$, located in a homogeneous thin plate. The real and imaginary parts of this complex valued function are plotted in logarithmic scale in Figs. 2(a) and 2(b) for $\beta_p a = 0.2$ and for different values of the loss factor $\gamma_b$. Particular patterns can be noticed due to the higher order dependence on the frequency (order 4). An equivalent of the polarizability (in electromagnetism) can be derived for flexural waves, $\alpha_b$ for subwavelength inhomogeneities (that relates the response of the scatterer to the incident field). $\alpha_b$ can be expressed thus as

$$\alpha_b = \frac{\pi a^2}{8}(D_{in}-1)\left[\frac{1}{D_{in}+f_0(\nu)} + \frac{1}{D_{in}+f_2(\nu)}\right]. \tag{14}$$

$\alpha_b$ has thus the unit of a surface since the scattering flexural problem is of two dimensional nature (in electromagnetics or acoustics it has the unit of a volume, for example). Figures 2(c) and 2(d) show the absolute value (in logarithmic scale) and phase of $\alpha_b$ versus the normalized wavenumber. It can be immediately seen that similarly to Fig. 1, $\alpha_b$ experiences a resonant enhancement when the denominators $|D_{in}(\beta_0)+f_0(\nu)|$ or $|D_{in}(\beta_0)+f_2(\nu)|$ in Eq. (14) are minimum. The resonance criterion is thus met at two wavenumbers $\beta_p/[1+f_0(\nu)]$ and $\beta_p/[1+f_2(\nu)]$, showing also that these resonances are very sensitive to the environment. The enhancement is also higher for lower loss factor [Fig. 2(c)], since otherwise, the denominator would not completely vanish due to $\text{Im}[D_{in}] \neq 0$. The phase jump of $\pi$ around the resonance wavenumbers, shown in Fig. 2(d), confirms the resonant features of the flexural object.

**Set up of the scattering problem of corrugated structures.** Let us consider now a different structure, i.e. a corrugated cylinder shown in Fig. 3(a). This structure consists of an inner cylinder of radius $a$, and outer cylinder of radius $a_c$ and subwavelength periodic grooves of angle $\theta_N = (2\pi)/(2N) = \pi/N$. The material filling the grooves is considered to be the same as the material of the surrounding (thin plate) and the boundary conditions at $r = a$ and at the boundary of the groves are either rigid (clamped) or stress-free, as given in the previous section. Inside the grooves, using the fact that $\beta_0(\pi a_c / N) \ll 1$, i.e. subwavelength corrugations (the period of the corrugations is $\pi a_c / N$), it is well known that only the fundamental waveguide mode is non negligible [14]. This leads to the following expression of the displacement field for $a \leq r \leq a_c$,

$$\zeta^{in}(r,\phi) = \sum_{l=0}^{\infty} \varepsilon_l i^l \left[ C_l Y_0(\beta_0 r) + C_l' K_0(\beta_0 r) + E_l J_0(\beta_0 r) + F_l I_0(\beta_0 r) \right] \cos l\phi. \quad (15)$$

For the internal boundary $r = a$, clamped boundary conditions shall be used first, that is $\zeta = \partial \zeta / \partial r = 0$. Applying the six boundary conditions for the system depicted in Fig. 3(a), with the remaining fields in other domains having similar expressions as in the previous section, for each azimuthal order $l$, we obtain a matrix system of equations. In particular, for the scattering unknowns $A_l = \psi_l / \chi_l$ and $B_l = \xi_l / \chi_l$:

$$\psi_l = \begin{vmatrix} -J_l(\beta_0 a_c) & K_l(\beta_0 a_c) & -Y_0(\beta_0 a_c) & -K_0(\beta_0 a_c) & -J_0(\beta_0 a_c) & -I_0(\beta_0 a_c) \\ 0 & 0 & Y_0(\beta_0 a) & K_0(\beta_0 a) & J_0(\beta_0 a) & I_0(\beta_0 a) \\ -\beta_0 J_l'(\beta_0 a_c) & \beta_0 K_l'(\beta_0 a_c) & -\beta_0 Y_0'(\beta_0 a_c) & -\beta_0 K_0'(\beta_0 a_c) & -\beta_0 J_0'(\beta_0 a_c) & -\beta_0 I_0'(\beta_0 a_c) \\ 0 & 0 & \beta_0 Y_0'(\beta_0 a) & \beta_0 K_0'(\beta_0 a) & \beta_0 J_0'(\beta_0 a) & \beta_0 I_0'(\beta_0 a) \\ -S_J(\beta_0 a_c) & S_K(\beta_0 a_c) & -\delta S_Y(\beta_0 a_c) & -\delta S_K(\beta_0 a_c) & \delta S_J(\beta_0 a_c) & \delta S_I(\beta_0 a_c) \\ -T_J(\beta_0 a_c) & T_K(\beta_0 a_c) & -\delta T_Y(\beta_0 a_c) & -\delta T_K(\beta_0 a_c) & \delta T_J(\beta_0 a_c) & \delta T_I(\beta_0 a_c) \end{vmatrix}, \quad (16)$$

$$\chi_l = \begin{vmatrix} H_l^{(1)}(\beta_0 a_c) & K_l(\beta_0 a_c) & -Y_0(\beta_0 a_c) & -K_0(\beta_0 a_c) & -J_0(\beta_0 a_c) & -I_0(\beta_0 a_c) \\ 0 & 0 & Y_0(\beta_0 a) & K_0(\beta_0 a) & J_0(\beta_0 a) & I_0(\beta_0 a) \\ \beta_0 H_l^{(1)'}(\beta_0 a_c) & \beta_0 K_l'(\beta_0 a_c) & -\beta_0 Y_0'(\beta_0 a_c) & -\beta_0 K_0'(\beta_0 a_c) & -\beta_0 J_0'(\beta_0 a_c) & -\beta_0 I_0'(\beta_0 a_c) \\ 0 & 0 & \beta_0 Y_0'(\beta_0 a) & \beta_0 K_0'(\beta_0 a) & \beta_0 J_0'(\beta_0 a) & \beta_0 I_0'(\beta_0 a) \\ S_H(\beta_0 a_c) & S_K(\beta_0 a_c) & -\delta S_Y(\beta_0 a_c) & -\delta S_K(\beta_0 a_c) & \delta S_J(\beta_0 a_c) & \delta S_I(\beta_0 a_c) \\ T_H(\beta_0 a_c) & T_K(\beta_0 a_c) & -\delta T_Y(\beta_0 a_c) & -\delta T_K(\beta_0 a_c) & \delta T_J(\beta_0 a_c) & \delta T_I(\beta_0 a_c) \end{vmatrix}, \quad (17)$$

and

$$\xi_l = \begin{vmatrix} H_l^{(1)}(\beta_0 a_c) & -J_l(\beta_0 a_c) & -Y_0(\beta_0 a_c) & -K_0(\beta_0 a_c) & -J_0(\beta_0 a_c) & -I_0(\beta_0 a_c) \\ 0 & 0 & Y_0(\beta_0 a) & K_0(\beta_0 a) & J_0(\beta_0 a) & I_0(\beta_0 a) \\ \beta_0 H_l^{(1)\prime}(\beta_0 a_c) & -\beta_0 J_l'(\beta_0 a_c) & -\beta_0 Y_0'(\beta_0 a_c) & -\beta_0 K_0'(\beta_0 a_c) & -\beta_0 J_0'(\beta_0 a_c) & -\beta_0 I_0'(\beta_0 a_c) \\ 0 & 0 & \beta_0 Y_0'(\beta_0 a) & \beta_0 K_0'(\beta_0 a) & \beta_0 J_0'(\beta_0 a) & \beta_0 I_0'(\beta_0 a) \\ S_H(\beta_0 a_c) & -S_J(\beta_0 a_c) & -\delta S_Y(\beta_0 a_c) & -\delta S_K(\beta_0 a_c) & \delta S_J(\beta_0 a_c) & \delta S_I(\beta_0 a_c) \\ T_H(\beta_0 a_c) & -T_J(\beta_0 a_c) & -\delta T_Y(\beta_0 a_c) & -\delta T_K(\beta_0 a_c) & \delta T_J(\beta_0 a_c) & \delta T_I(\beta_0 a_c) \end{vmatrix}, \quad (18)$$

with the same notations for the functionals $S_Z$ and $T_Z$ given in Eq. (9), with the main difference here, is that for columns 3-6, only Bessel or Hankel functions of order $l = 0$ are taken into account, due to the subwavelength nature of the corrugations [Eq. (15)] and the presence of the extra term $\delta = \theta_1 / (\theta_1 + \theta_2)$, accounting for the filling factor of the corrugations ($\delta = 0.5$ if $\theta_1 = \theta_2$). The other possibility with flexural waves is to ensure stress-free boundary conditions at $r = a$, that is $M_r = V_r = 0$ [given in Eqs. (6)-(7)]. Applying the six boundary conditions for the system, for each azimuthal order $l$, we obtain again an algebraic system of equations, in particular for the scattering unknowns $A_l = \psi_l / \chi_l$ and $B_l = \xi_l / \chi_l$ (here we keep the same notations for the scattering coefficients for simplicity):

$$\psi_l = \begin{vmatrix} -J_l(\beta_0 a_c) & K_l(\beta_0 a_c) & -Y_0(\beta_0 a_c) & -K_0(\beta_0 a_c) & -J_0(\beta_0 a_c) & -I_0(\beta_0 a_c) \\ -\beta_0 J_l'(\beta_0 a_c) & \beta_0 K_l'(\beta_0 a_c) & -\beta_0 Y_0'(\beta_0 a_c) & -\beta_0 K_0'(\beta_0 a_c) & -\beta_0 J_0'(\beta_0 a_c) & -\beta_0 I_0'(\beta_0 a_c) \\ -S_J(\beta_0 a_c) & S_K(\beta_0 a_c) & -\delta S_Y(\beta_0 a_c) & -\delta S_K(\beta_0 a_c) & \delta S_J(\beta_0 a_c) & \delta S_I(\beta_0 a_c) \\ 0 & 0 & \delta S_Y(\beta_0 a) & \delta S_K(\beta_0 a) & \delta S_J(\beta_0 a) & \delta S_I(\beta_0 a) \\ -T_J(\beta_0 a_c) & S_K(\beta_0 a_c) & -\delta T_Y(\beta_0 a_c) & -\delta T_K(\beta_0 a_c) & \delta T_J(\beta_0 a_c) & \delta T_I(\beta_0 a_c) \\ 0 & 0 & \delta T_Y(\beta_0 a) & \delta T_K(\beta_0 a) & \delta T_J(\beta_0 a) & \delta T_I(\beta_0 a) \end{vmatrix}, \quad (19)$$

$$\chi_l = \begin{vmatrix} H_l^{(1)}(\beta_0 a_c) & K_l(\beta_0 a_c) & -Y_0(\beta_0 a_c) & -K_0(\beta_0 a_c) & -J_0(\beta_0 a_c) & -I_0(\beta_0 a_c) \\ \beta_0 H_l^{(1)'}(\beta_0 a_c) & \beta_0 K_l'(\beta_0 a_c) & -\beta_0 Y_0'(\beta_0 a_c) & -\beta_0 K_0'(\beta_0 a_c) & -\beta_0 J_0'(\beta_0 a_c) & -\beta_0 I_0'(\beta_0 a_c) \\ S_H(\beta_0 a_c) & S_K(\beta_0 a_c) & -\delta S_Y(\beta_0 a_c) & -\delta S_K(\beta_0 a_c) & \delta S_J(\beta_0 a_c) & \delta S_I(\beta_0 a_c) \\ 0 & 0 & \delta S_Y(\beta_0 a) & \delta S_K(\beta_0 a) & \delta S_J(\beta_0 a) & \delta S_I(\beta_0 a) \\ T_H(\beta_0 a_c) & S_K(\beta_0 a_c) & -\delta T_Y(\beta_0 a_c) & -\delta T_K(\beta_0 a_c) & \delta T_J(\beta_0 a_c) & \delta T_I(\beta_0 a_c) \\ 0 & 0 & \delta T_Y(\beta_0 a) & \delta T_K(\beta_0 a) & \delta T_J(\beta_0 a) & \delta T_I(\beta_0 a) \end{vmatrix}, \quad (20)$$

and

$$\xi_l = \begin{vmatrix} H_l^{(1)}(\beta_0 a_c) & -J_l(\beta_0 a_c) & -Y_0(\beta_0 a_c) & -K_0(\beta_0 a_c) & -J_0(\beta_0 a_c) & -I_0(\beta_0 a_c) \\ \beta_0 H_l^{(1)'}(\beta_0 a_c) & -\beta_0 J_l'(\beta_0 a_c) & -\beta_0 Y_0'(\beta_0 a_c) & -\beta_0 K_0'(\beta_0 a_c) & -\beta_0 J_0'(\beta_0 a_c) & -\beta_0 I_0'(\beta_0 a_c) \\ S_H(\beta_0 a_c) & -S_J(\beta_0 a_c) & -\delta S_Y(\beta_0 a_c) & -\delta S_K(\beta_0 a_c) & \delta S_J(\beta_0 a_c) & \delta S_I(\beta_0 a_c) \\ 0 & 0 & \delta S_Y(\beta_0 a) & \delta S_K(\beta_0 a) & \delta S_J(\beta_0 a) & \delta S_I(\beta_0 a) \\ T_H(\beta_0 a_c) & -S_J(\beta_0 a_c) & -\delta T_Y(\beta_0 a_c) & -\delta T_K(\beta_0 a_c) & \delta T_J(\beta_0 a_c) & \delta T_I(\beta_0 a_c) \\ 0 & 0 & \delta T_Y(\beta_0 a) & \delta T_K(\beta_0 a) & \delta T_J(\beta_0 a) & \delta T_I(\beta_0 a) \end{vmatrix}. \quad (21)$$

**Localized SPMs induced on corrugated thin plate cylinders.** Consider first the case of corrugations with clamped boundary conditions, shown in Fig. 3(a) with its equivalent platonic metamaterial (effective flexural rigidity $D_{\text{eff}}$ and effective density $\rho_{\text{eff}}$) in Fig. 3(b), and whose scattering coefficients are deduced from Eqs. (16)-(18). The normalized scattering cross section $\sigma^{\text{scat}}/a$ is plotted versus the normalized flexural wavenumber $\beta_0 a$ for the range 0.25 to 3 in Fig. 3(b).

As known in the theory of scattering of thin plates [33]-[35],[37], rigid objects, i.e. described with boundary conditions $\zeta = \partial \zeta / \partial r = 0$ at their outer boundary, possess a divergent scattering cross section at zero frequency. This means that pins in thin plates have a natural resonance in the quasistatic regime. This is confirmed in the plot given in the inset of Fig. 3(b), where the SCS of Eq. (12) is plotted against the normalized wavenumber and extremely high values of normalized SCS can be seen. The purpose of our study of SPM is to demonstrate flexural resonances at higher frequencies that can be useful for a wide range of applications including sensing, imaging or camouflaging. Since there is no absorption from the structure (no loss from neither the thin plate nor the rigid scatterer), the SCS is equal to the overall extinction. The analysis of the spectral dependence of the normalized SCS, i.e. $\sigma^{scat}/a$ (with respect to the normalized wavelength $\beta_0 a$) shows the existence, in addition to the zero-frequency resonance (associated with any rigid obstacle), of several resonance peaks corresponding to the different resonances of the quasi-grating structure, described by the term of Eq. (15). In fact, as in the electromagnetic or acoustic scenarios [24],[5], one can interpret the corrugated cylinder as an equivalent medium with effective parameters that obey a *Drude*-like dispersion relation [27]. In fact, this analogy between plasmonics and platonics makes sense, since we have seen in the first section, that the condition of resonant scattering from a cylinder is a negative flexural density. The formula of Eq. (13) permits these negative values and is therefore a good description of the behavior of the SCS of Fig. 3(b). By carefully observing the expressions of the scattering coefficients of Eqs. (16)-(18), major differences with the scattering of core-shells structures can be observed. To put it clear, the scattering from a rigid (or stress-free) cylinder coated with a shell is also described by a $6 \times 6$ determinants. The main difference is the form of the displacement field $\zeta$ in the shell, i.e. only the zero order Bessel and Hankel functions is taken into account in Eq. (15), due to the subwavelength nature of the corrugations. The second difference is the presence of an extra term in the determinants, i.e. $\delta = \theta_1/(\theta_1 + \theta_2)$ that accounts for the filling fraction of the grooves. These two differences make the SCS of the corrugated structure very different and much richer than the scattering of classical core-shell structures [34]-[35],[37],[38]. The SCS of Fig. 3(b) confirms these predictions and several multipole resonances of this rather simple structure can be observed for different wavenumbers. At very low wavenumbers, the resonance associated with the rigid object is somewhat of greater amplitude since it is inversely proportional to $\beta_0$ as defined in Eq. (12). The resonance peaks locations and bandwidths are also very sensitive to the filling factor of the grooves $\delta$. Two values are chosen in Fig. 3(b), i.e. 0.1 (dilute regime) and 0.7 (high filling) and they show that for higher filling, one can obtain broader resonances. The low filling results also in a blueshift (towards higher frequencies) of all the multipole resonances. It also remarkable to note that the higher order multipoles (at higher frequencies) are more sensitive to loss since they have a very narrowband. One can thus expect that these modes can be efficiently used for sensing due to their extreme sensitivity to the environment. However, one must note also that if high amounts of loss of the structure are present, some of these modes may simply vanish and only the broadband mode (at lower frequencies) may remain.

What is unique about the scattering theory in thin plates, as already stressed, is the possibility to have various boundary conditions. The second type of boundary conditions that can be applied to the structure of Fig. 3(a) is the stress-free condition [given by $M_r = V_r = 0$, in Eqs. (6)-(7)]. The corresponding scattering coefficients of this scattering problem are given by Eqs. (19)-(21), clearly very different from those of Eqs. (16)-(18). The main difference with the rigid case, is that for a uniform cylinder, the SCS vanishes at low frequencies [no singularity as the one observed in the inset of Fig. 3(b)]. So the scattering by stress-free obstacles is by definition non-resonant. However, inducing corrugations in the structure, as schematized in Fig. 3(a), results in multipolar resonances as for the rigid case. This can be verified in Fig. 4(a), where the SCS is plotted for geometry and parameters similar to the ones of Fig. 3. One can see that the normalized SCS undergoes in the same spectral range successive resonances of different amplitudes and bandwidths. To verify that these peaks are associated to the multipolar resonances of the corrugated structure, the different (normalized) multipolar coefficients $4/(\beta_0 a)|A_l|^2$ are also given with different colors in the same Fig. 4(a). These plots show that each resonant peak of the SCS can be associated with a resonance of a specific scattering coefficient with given order $l$. The zero-scattering order (red line) has incidentally no resonant enhancement for the parameters and geometry of the proposed structure.

In order to get more insight about the nature of the different resonances seen in Fig. 4(a), Fig. 4(c) plots the amplitude of the displacement field distribution $\zeta$ around specific normalized wavenumbers of the corresponding modes. From this figure, the nature of the resonant modes can be clear seen.

**Dispersion relation of the SPMs.** In this section, we would like to analyze the tunability properties of the corrugated structure of Fig. 3(a) as well as the possibility to treat it as an effective metamaterial with homogeneous flexural rigidity as schematized in Fig. 3(a). In order to better understand the behavior of the corrugated cylinder, the dispersion curves of the SPM originating from the one-dimensional equivalent grating of same geometrical parameters. The resonant wavenumber of the equivalent material is taken as $\beta_p a = \pi/\sqrt{2}$. The normalized frequency $(\beta_0 a)^2$ is given versus the normalized propagation constant of the normalized platonic spoof plasmon $\beta_{SPM} a$ in Fig. 5(a) for different loss factors. It should be noted here that the dispersion of waves in free space is of parabolic nature and is not linear as is the case for acoustics and electromagnetism. This means that elastic cone is of parabolic shape (red curve). It could be seen that for lower wavenumbers, the dispersion relation is identical to the free space one. For higher wave numbers, the behavior changes and the dispersion of the SPM becomes flat (with convergence to $\sqrt{2}\beta_p$) and located outside the elastic cone. This behavior is characteristic of localized surface plasmons polaritons in optical frequencies and shows undoubtedly that the corrugated acoustic cylinder of Fig. 3(a) gives rise to similar features in elasticity (flexural waves in thin plates), which is unprecedented to the best of the knowledge of the authors. Additionally from the dispersion relations of Fig. 5(b), one can conclude that ultimately, the dispersion of SPMs is very sensitive to the environment parameters and therefore can be a platform for elasticity sensors.

**Summary.** We have introduced here the concept and potential realization of an elastic localized surface mode, obtained by a subwavelength corrugation of a rigid or stress-free elastic cylinder in a thin plate. We have demonstrated with analytical analysis that the features of these plasmons are very much similar to their electromagnetic and acoustic counterpart. The experimental realization of this idea [structure of Fig. 3(a)] may be within reach in the near future (note for instance that lensing of bending waves via negative refraction was theoretically predicted using the biharmonic plate model [36] and experimentally confirmed in a thin Duralumium plate [13]), allowing for exciting applications of interest to elastodynamics, including subwavelength imaging and sensing for the oil and gas industry, elastic wave guiding, and enhancement of nonlinear effects

**Methods.** Analytical methods based on scattering Mie theory of cylindrical objects in thin elastic plates are used to obtain the numerical simulations in this study. The vertical displacement of the plate is the solution of the fourth order differential biharmonic equation. We proceed, as usually done, by expanding the impinging plane waves and the scattered fields in terms of Bessel and Hankel functions in polar coordinate system centered with the object. We then apply convenient elastodynamic boundary conditions on each cylindrical interface in order to obtain the scattering coefficients for waves, which uniquely determine the displacement fields everywhere. The displacement field distributions and scattering cross sections are computed using Bessel developments and Eq. (12) respectively. In the quasi-static limit, where the size of the elastic core sphere is much smaller than the wavelength and only the lowest-order Mie coefficient is important, an analytical formula is obtained [Eq. (10)]. Proper convergence for all the results is reached.

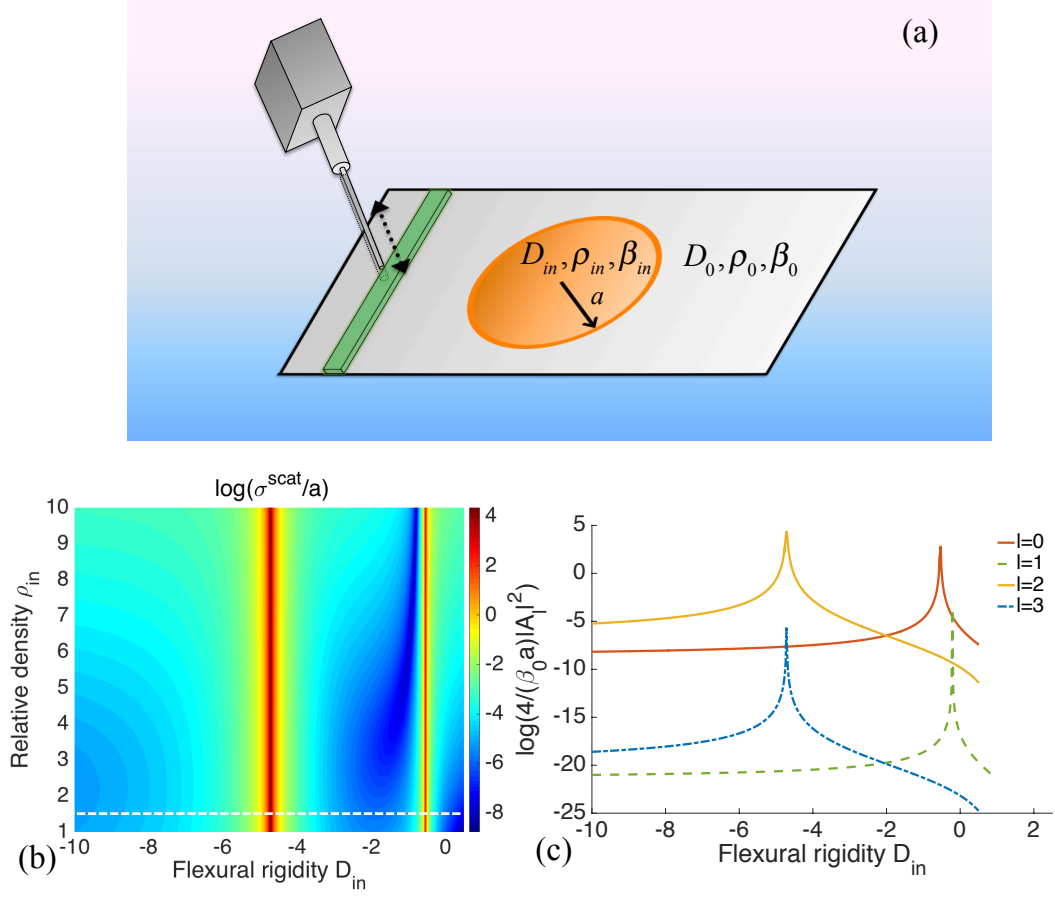

FIG. 1. (Color online) (a) Sketch of the scattering problem. (b) Normalized scattering cross section $\sigma^{scat}/a$ in logarithmic scale, of the cylinder versus its relative density $\rho_{in}$ and flexural rigidity $D_{in}$ for the normalized flexural wavenumber $\beta_0 a = 0.1$ and a Poisson ratio $\nu = 0.3$. (c) Normalized scattering multipoles ($l = 0,1,2,3$) of the cylindrical inhomogeneity in logarithmic scale for the same wavenumber and for $\rho_{in} = 1.5$, showing that the fundamental ($l = 0$) and second order ($l = 2$) multipole are orders of magnitude higher than the remaining ones.

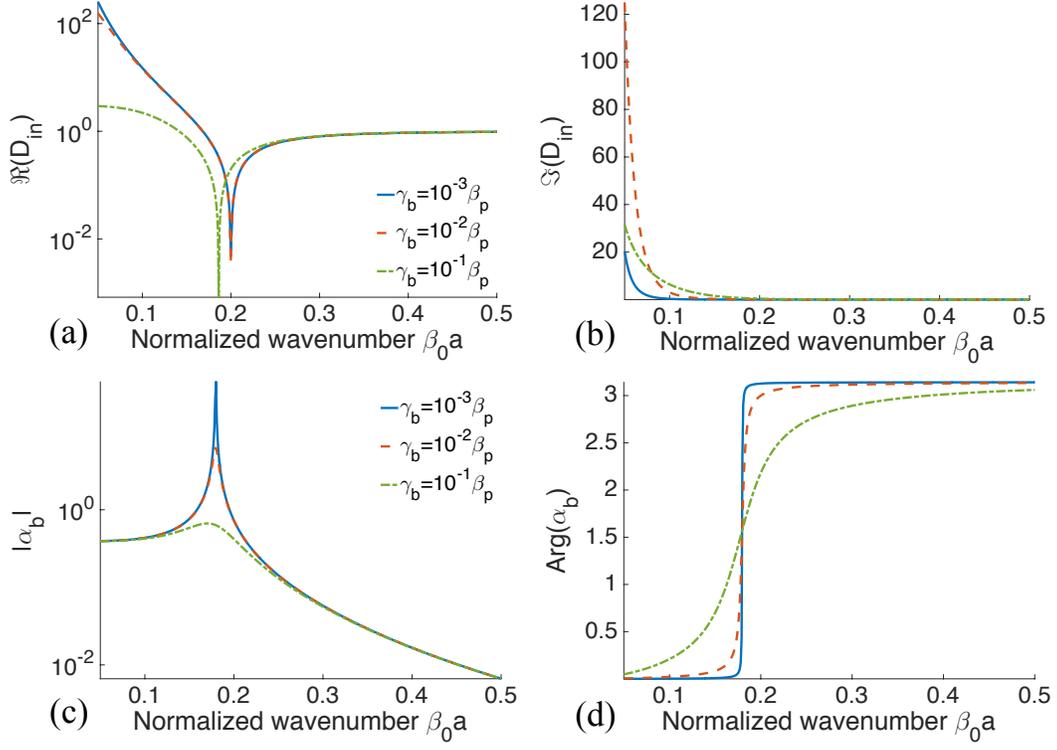

FIG. 2. (Color online) (a) Real part of the dispersive flexural rigidity $D_{in}$ in logarithmic scale. (b) Imaginary part of $D_{in}$. (c) Amplitude of the equivalent of polarizability $\alpha_b$ in logarithmic scale for the corresponding flexural rigidity given in Fig. 2(a) and for density $\rho_{in}=1$. (d) Phase of $\alpha_b$ for the same parameters.

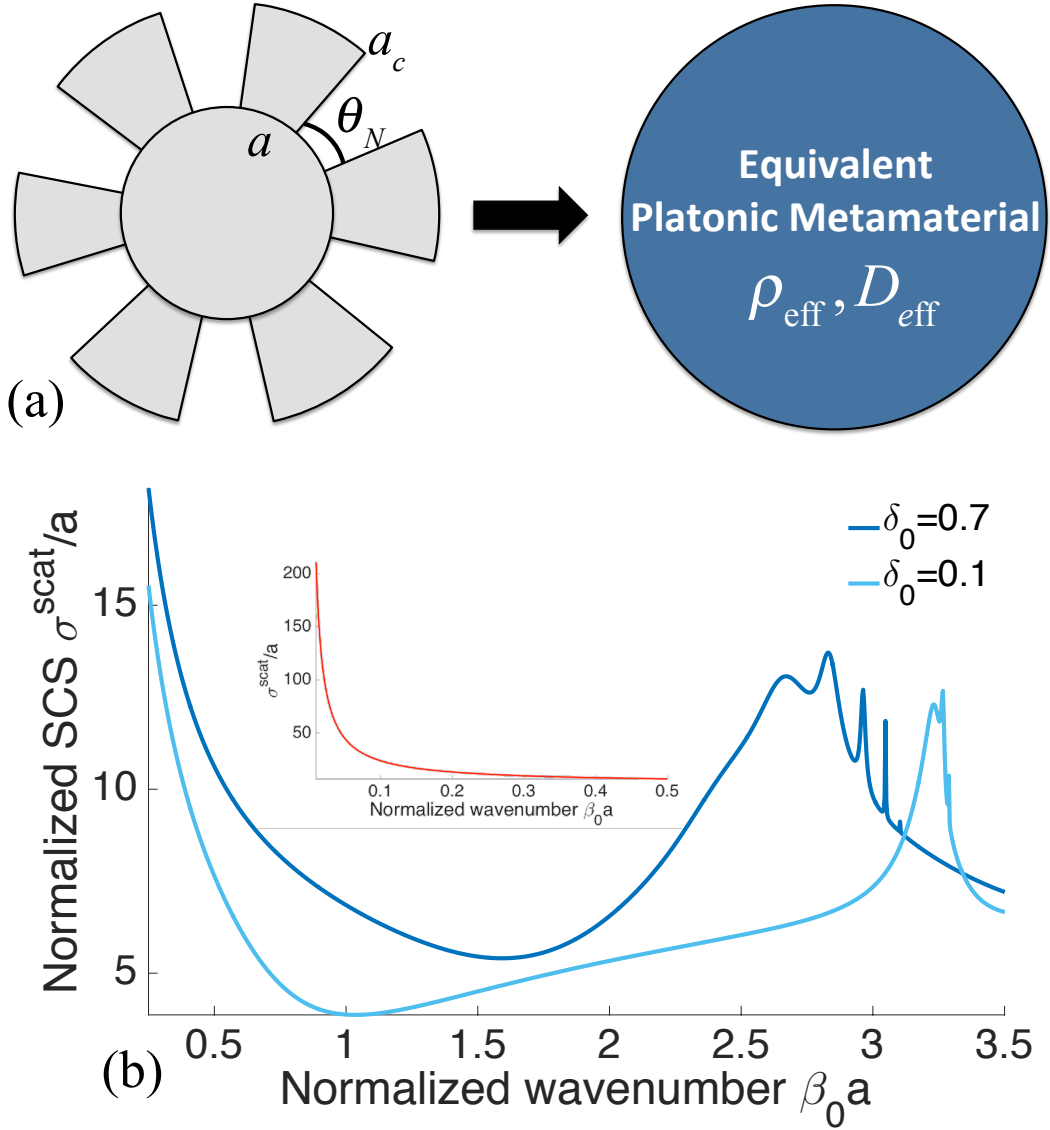

FIG. 3. (Color online) (a) Equivalent platonic metamaterial model for the corrugated elastic cylinder (with grooves of angle $\theta_N = \pi/N$ and its effective density $\rho_{eff}$ and flexural rigidity $D_{eff}$. (b) Normalized scattering cross section $\sigma^{scat}/a$ of the structure shown in (a), versus the normalized wavenumber. The inset in (b) shows the scattering cross section of a rigid cylinder of same radius showing resonant scattering for $\beta_0 a \approx 0$ and no resonant effects for higher frequencies.

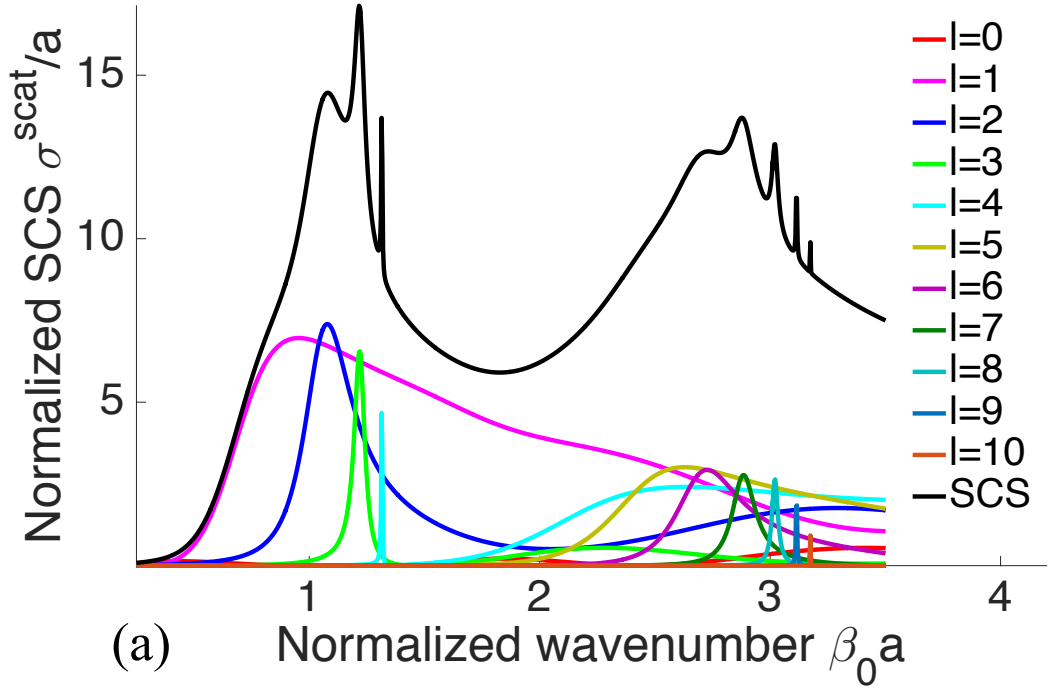

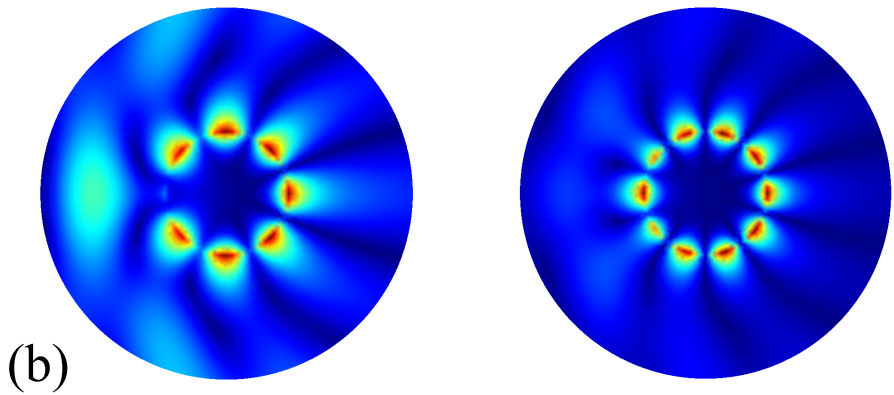

FIG. 4. (Color online) (a) Normalized scattering cross section $\sigma^{scat}/a$ of the corrugated cylinder shown in Fig. 2(a) versus the normalized wavenumber, using stress-free boundary conditions (black curve). The different multipoles ($l = 0...10$) corresponding to determinants in Eqs. (18)-(20) are plotted to show the contribution of each multipolar resonance. (b) Near field plots of the displacement $\zeta(r,\phi)$ for normalized wavenumbers 1.3 and 2.75, corresponding to modes $l = 4$ and $l = 6$.

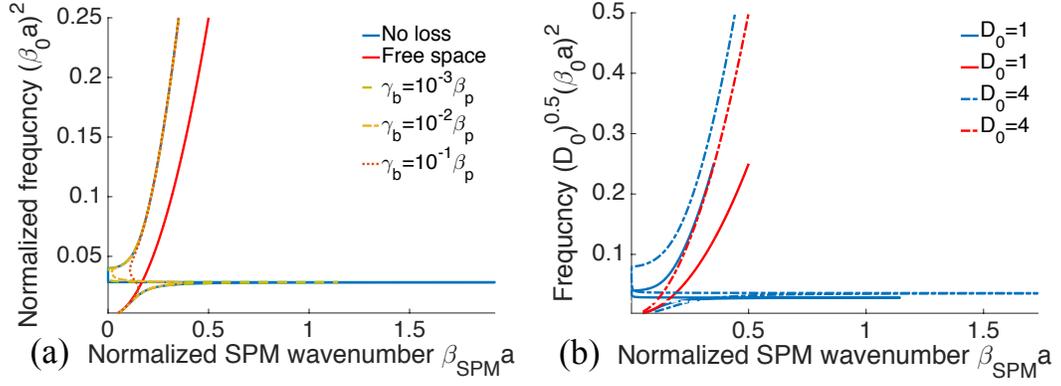

FIG. 5. (Color online) (a) Dispersion relation of the SPM versus normalized frequency $(\beta_0 a)^2$ for $D_0 = 1$ and for different values of the loss factor $\gamma_b = \{0, 10^{-3}, 10^{-2}, 10^{-1}\} \times \beta_p$, with $\beta_p$ the natural wavenumber of the model of Eq. (13). (b) Same as (a) but for different values of the relative flexural rigidity of the surrounding medium $D_0 = 1$ and $4$.